\documentclass{pasj00}

\begin{document}
\SetRunningHead{Y. Soejima et al.}{Three Superoutbursts of SW UMa}
\Received{}
\Accepted{}

\title{Photometry of Three Superoutbursts of the SU
UMa-type Dwarf Nova, SW Ursae Majoris}

\author{
Yuichi \textsc{Soejima}, \altaffilmark{1*}
Daisaku \textsc{Nogami}, \altaffilmark{2}
Taichi \textsc{Kato}, \altaffilmark{1}
Makoto \textsc{Uemura}, \altaffilmark{3}
Akira \textsc{Imada}, \altaffilmark{4}
\\
Kei \textsc{Sugiyasu}, \altaffilmark{1}
Hiroyuki \textsc{Maehara}, \altaffilmark{2}
Ken'ichi \textsc{Torii}, \altaffilmark{5}
Kenji \textsc{Tanabe}, \altaffilmark{6}
Arto \textsc{Oksanen}, \altaffilmark{7}
\\
Kazuhiro \textsc{Nakajima}, \altaffilmark{8}
Rudolf \textsc{Nov\'{a}k}, \altaffilmark{9}
Gianluca \textsc{Masi}, \altaffilmark{10}
Tom\'{a}\v{s} \textsc{Hynek}, \altaffilmark{11}
Brian \textsc{Martin}, \altaffilmark{12}
\\
Denis \textsc{Buczynski}, \altaffilmark{13}
Elena P. \textsc{Pavlenko}, \altaffilmark{14}
Sergei Yu. \textsc{Shugarov}, \altaffilmark{15}
Lewis M. \textsc{Cook}, \altaffilmark{16}
}

\altaffiltext{1}{Department of Astronomy, Faculty of Science, Kyoto
University, Sakyo-ku, Kyoto 606-8501} 
\email{$^{*}$soejima@kusastro.kyoto-u.ac.jp}

\altaffiltext{2}{Kwasan Observatory, Kyoto University, Yamashina-ku,
Kyoto 607-8471}

\altaffiltext{3}{Astrophysical Science Center, Hiroshima University,
Kagamiyama, 1-3-1 Higashi-Hiroshima 739-8526}

\altaffiltext{4}{Space Research Course in the Department of Physics,
Faculty of Science, \\ Kagoshima University, Korimoto, Kagoshima 890-0065}

\altaffiltext{5}{Department of Earth and Space Science, Graduate School of
Science, \\ Osaka University, 1-1 Machikaneyama-cho, Toyonaka, Osaka 560-0043}

\altaffiltext{6}{Department of Biosphere-Geosphere System Science, Faculty of
Informatics, Okayama University of Science, \\ 1-1 Ridai-cho, Okayama, Okayama 700-0005}

\altaffiltext{7}{Nyrola Observatory, Jyvaskylan Sirius ry, Kyllikinkatu
1, FIN-40100 Jyvaskyla, Finland}

\altaffiltext{8}{VSOLJ, 124 Isatotyo, Teradani, Kumano, Mie 519-4673}

\altaffiltext{9}{Institute of Computer Science, Faculty of Civil
 Engineering, Brno University of Technology, 602 00 Brno}

\altaffiltext{10}{The Virtual Telescope Project, Via Madonna del Loco 47, 03023
Ceccano (FR), Italy}

\altaffiltext{11}{Observatory and Planetarium of Johann Palisa, VSB -- Technical 
\\ University Ostrava, Trida 17. listopadu 15, Ostrava -- Poruba 708 33, 
Czech Republic}

\altaffiltext{12}{The King's University College; Center for Backyard Astrophysics 
\\ (Alberta), Edmonton, Alberta, Canada T6B 2H3}

\altaffiltext{13}{Conder Brow Observatory, Littlefell Lane, Lancaster
LA1 IXD, England}

\altaffiltext{14}{Crimean Astrophysical Observatory, 98409, Nauchny,
Crimea, Ukraine}

\altaffiltext{15}{Sternberg Astronomical institute, Moscow
University, Universitetsky Ave., 13, Moscow 119992, Russia, and
Astronomical 
\\ Institute of Slovak Academy of Sciences, 05960, Tatranska Lomnica, Slovakia}

\altaffiltext{16}{Center for Backyard Astrophysics (Concord), 1730 Helix
Ct. Concord, California 94518, USA}


\maketitle

\begin{abstract}

We investigated the superhump evolution, analysing optical photometric
observations of the 2000 February-March, the 2002 October-November, and
the 2006 September superoutbursts of SW UMa. 
The superhumps evolved in the same way after their appearance during
the 2000 and the 2002 superoutbursts, and probably during the 2006 one. 
This indicates that the superhump evolution may be governed by the
invariable binary parameters. We detected a periodicity in light
curve after the end of the 2000 superoutburst without phase shift,
which seems to be the remains of the superhumps.
We found QPOs at the end stage of the 2000 and the 2002 superoutbursts,
but failed to find extraordinarily large-amplitude QPOs called
`super-QPOs' which previously have been observed in SW UMa.

\end{abstract}

\section{INTRODUCTION}
\label{introduction}

Cataclysmic variables (CVs) are close binaries containing a white dwarf
(primary) and a late-type star (secondary).
The secondary fills its Roche-lobe, and 
transfers gas to the primary, so that if the primary star is
non-magnetic, an accretion disk is formed from the material
spiraling onto the white dwarf (for a review, see
e.g. \cite{war95book}; \cite{hel01book}; \cite{smi07review}). 

Dwarf novae are a subclass of CVs, and undergo
outbursts which are caused by the thermal instability in the accretion disk
\citep{osa74DNmodel}. SU UMa-type dwarf novae are the most spectacular
subgroup of the dwarf novae which are characterized by two distinct
types of outbursts: more frequent normal outbursts lasting typically for
a few days, and less frequent long and large-amplitude superoutbursts
lasting for about two weeks. The most enigmatic feature of the
superoutbursts is an occurrence of photometric light humps with an
amplitude of about 0.2-0.3 mag, called ``superhumps'' which repeat
with a few percent longer period than that of the orbital motion.

\citet{osa89suuma} has explained the general behavior of
SU UMa stars by combining the thermal instability and the tidal
instability. It is thus called the thermal-tidal instability model.
According to this model, the disk radius is expanded at the beginning of
the outburst by the increased viscosity. When an outburst pushes the
outer disk beyond the critical radius for the 3:1 resonance, the tidal
instability is triggered, producing a precessing eccentric disk.
Superhumps can be explained as a beat phenomenon of the precession of
the tidally deformed disk and the orbital motion of the system.

SW UMa is an SU UMa-type dwarf nova with a short orbital period of 81.8 min
\citep{sha86swumaXray}, and various observations of this object were
carried out so far. During the 1986 superoutburst, its superhumps were detected
by \citet{rob87swumaQPO} for the first time, and they determined the
superhump period ($P_{SH}$) as to be 84.0 min (0.5833 days). 
\citet{kat92swumasuperQPO} found `super-QPOs'
with an extraordinarily large amplitude of $\sim0.2$ mag and a period of
$\sim6$ min during the 1992 superoutburst, which is a peculiar
phenomenon of the superoutbursts of SW UMa.
\citet{sem97swuma} found that the $P_{SH}$ was 0.05818(2) days, and the
$P_{SH}$ derivative ($P_{dot}=\dot{P}_{SH}/P_{SH}$) was $8.9 \times
10^{-5}$ during the 1996 superoutburst. \citet{nog98swuma}
also detected $P_{SH}=0.05818(2)$ days, and $P_{dot}=8.8(0.7) \times
10^{-5}$ during the same
superoutburst. \citet{nog98swuma} found QPOs, but failed to
detect `super-QPOs'.
In quiescence, SW UMa is at $V=$16.5-17. A 15.9-min periodicity was
discovered in its optical light curve, and
also marginally detected in the soft X-ray data by $EXOSAT$
\citep{sha86swumaXray}. This
indicates that SW UMa has the nature of an intermediate polar
harboring a strongly magnetized white dwarf rotating with a period
of 15.9 min. 
Recently, \citet{pav00swuma} observed the late stage of the 2000 superoutburst of
SW UMa, and found late superhumps with a period of 0.1197 days and the
15.9-min oscillations.

In this paper, we analysed the photometric observations of SW UMa during
the 2000, 2002, and 2006 superoutbursts. The results
of the analyses are summarized in the section \ref{result}. We will
discuss the properties in light curves, the superhump evolution, and
QPOs in the section
\ref{discussion}. Our conclusions are put in the last section \ref{conclusion}.

\section{OBSERVATION}

We analysed the data of time-resolved CCD photmetries by VSNET
Collaborators. The summary of the observations is given in table \ref{obs1},
\ref{obs2}, \ref{obs3}, and \ref{obs4}. We also used the data from the AAVSO
International Database and visual observations reported to VSNET for the
supplement. Heliocentric corrections to
the observation times, and correction for systematic differences between
observers were applied before the following analysis.

\begin{table*}
\caption{Observers and instruments.}
\label{obs1}
\begin{center}
\begin{tabular}{cllll}  
\hline \hline
Symbol$^{*}$ & Telescope & CCD & Observer & Site \\ 
\hline

A & 30 cm & ST-7 & Kyoto Team & Japan \\
B & 12.5 in & CB245 & B. Martin & Canada \\
C & 38 cm & SXLB & D. Buczynski & UK \\
D & 44 cm & CB245 & L. Cook & USA \\
E & 40 cm & ST-7 & R. Nov\'{a}k & Czech \\
F & 38 cm & ST-7 & E. Pavlenko & Ukraine \\
G & 30 cm & electrophotometer & S. Shugarov & Russia \\
H & 28 cm & ST-7 & G. Masi & Italy \\
I & 30 cm & ST-9 & K. Tanabe & Japan \\
J & 25 cm & AP6E & K. Torii & Japan \\
K & 20 cm & AP7p & K. Torii & Japan \\
L & 9 cm & ST-7 & T. Hynek & Czech \\
M & 40 cm & STL-1001E & A. Oksanen & Finland \\
N & 25 cm & ST-7XME & H. Maehara & Japan \\
O & 25 cm & CV04 & K. Nakajima & Japan \\
P & 40 cm & ST-7 & Kyoto Team & Japan \\

\hline 
\multicolumn{5}{l}{$^{\*}$ Symbols in table \ref{obs2} and \ref{obs3}.}\\ 
\end{tabular}
\end{center}
\end{table*}

\begin{table*}
\caption{Observations during the 2000 superoutburst.}
\label{obs2}  
\begin{center}
\begin{tabular}{lllclllc}
\hline\hline
Start date$^{*}$ & End date$^{*}$ &$N^{\dagger}$ & Observer$^{\ddagger}$
 & Start date$^{*}$ & End date$^{*}$ &$N^{\dagger}$ & Observer$^{\ddagger}$\\ 
\hline 

1586.950 & 1587.370 & 1261 & A & 1606.897 & 1607.148 & 218  & A \\
1587.899 & 1588.125 & 493  & A & 1607.258 & 1607.275 & 7    & G \\
1589.892 & 1590.079 & 803  & A & 1607.386 & 1607.502 & 200  & C \\
1590.894 & 1591.388 & 1445 & A & 1607.292 & 1607.417 & 160  & H \\
1591.616 & 1592.044 & 335  & B & 1608.283 & 1608.444 & 98   & F \\
1591.902 & 1592.382 & 1657 & A & 1608.490 & 1608.602 & 193  & C \\
1592.653 & 1592.902 & 200  & B & 1608.898 & 1609.344 & 726  & A \\
1592.901 & 1593.164 & 1109 & A & 1609.310 & 1609.532 & 79   & G \\
1593.685 & 1593.865 & 546  & D & 1609.899 & 1610.320 & 860  & A \\
1594.367 & 1594.477 & 215  & C & 1610.361 & 1610.440 & 40   & G \\
1594.900 & 1595.386 & 1524 & A & 1610.898 & 1611.345 & 1100 & A \\
1595.890 & 1596.372 & 1969 & A & 1611.902 & 1612.170 & 522  & A \\
1596.961 & 1597.379 & 1977 & A & 1612.318 & 1612.417 & 70   & F \\
1597.893 & 1598.171 & 1422 & A & 1612.981 & 1613.080 & 135  & A \\
1598.911 & 1599.382 & 936  & A & 1613.238 & 1613.267 & 17   & F \\
1599.890 & 1600.243 & 1091 & A & 1613.903 & 1614.101 & 442  & A \\
1601.020 & 1601.378 & 210  & A & 1614.403 & 1614.555 & 57   & G \\
1602.059 & 1602.369 & 1794 & A & 1615.216 & 1615.361 & 67   & F \\
1602.902 & 1603.298 & 983  & A & 1615.904 & 1616.127 & 552  & A \\
1602.405 & 1602.399 & 562  & E & 1616.905 & 1617.117 & 536  & A \\
1603.713 & 1603.967 & 194  & B & 1617.395 & 1617.519 & 63   & G \\
1604.406 & 1604.531 & 57   & G & 1617.902 & 1618.196 & 723  & A \\
1604.893 & 1605.278 & 2338 & A & 1618.227 & 1618.457 & 100  & F \\
1605.954 & 1606.355 & 995  & A & 1619.223 & 1619.446 & 80   & F \\
1606.279 & 1606.403 & 63   & F &&&& \\

\hline
\multicolumn{8}{l}{$^{*}$ HJD-2450000.} \\                           
\multicolumn{8}{l}{$^{\dagger}$ Number of frames.} \\                
\multicolumn{8}{l}{$^{\ddagger}$ See table \ref{obs1} for detail.}\\ 
\end{tabular}
\end{center}
\end{table*}

\begin{table*}
\caption{Observations during the 2002 superoutburst.}
\label{obs3}  
\begin{center}
\begin{tabular}{lllclllc}
\hline\hline
Start date$^{*}$ & End date$^{*}$ &$N^{\dagger}$ & Observer$^{\ddagger}$
 & Start date$^{*}$ & End date$^{*}$ &$N^{\dagger}$ & Observer$^{\ddagger}$\\ 
\hline 

2571.157 & 2571.244 & 425  & I & 2587.018 & 2587.325 & 556  & K \\
2572.182 & 2572.274 & 506  & I & 2587.194 & 2587.321 & 683  & I \\
2574.372 & 2574.504 & 93   & L & 2588.021 & 2588.333 & 684  & K \\
2575.040 & 2575.336 & 312  & J & 2588.193 & 2588.285 & 459  & I \\
2575.171 & 2575.347 & 1399 & I & 2589.038 & 2589.339 & 613  & K \\
2576.013 & 2576.339 & 634  & J & 2590.074 & 2590.396 & 339  & K \\
2576.188 & 2576.356 & 1029 & I & 2591.044 & 2591.347 & 372  & K \\
2577.075 & 2577.338 & 509  & J & 2591.301 & 2591.336 & 23   & I \\
2577.211 & 2577.354 & 697  & I & 2592.055 & 2592.344 & 575  & K \\
2578.067 & 2578.329 & 492  & J & 2594.313 & 2594.363 & 96   & I \\
2580.178 & 2580.295 & 476  & I & 2595.177 & 2595.360 & 980  & K \\
2580.187 & 2580.346 & 307  & J & 2595.289 & 2595.373 & 79   & I \\
2581.997 & 2582.337 & 647  & K & 2597.146 & 2597.376 & 500  & I \\
2582.247 & 2582.354 & 485  & I & 2598.146 & 2598.292 & 442  & K \\
2583.031 & 2583.351 & 700  & K & 2600.273 & 2600.379 & 131  & I \\
2583.284 & 2583.355 & 332  & I & 2601.258 & 2601.322 & 40   & I \\
2584.036 & 2584.339 & 652  & K & 2606.222 & 2606.270 & 101  & I \\
2584.219 & 2584.333 & 633  & I & 2607.198 & 2607.259 & 82   & I \\
2585.113 & 2585.328 & 343  & K & 2610.233 & 2610.256 & 40   & I \\
2586.137 & 2586.342 & 395  & K &&&& \\
				  
\hline
\multicolumn{8}{l}{$^{*}$ HJD-2450000.} \\                           
\multicolumn{8}{l}{$^{\dagger}$ Number of frames.} \\                
\multicolumn{8}{l}{$^{\ddagger}$ See table \ref{obs1} for detail.}\\ 
\end{tabular}
\end{center}
\end{table*}

\begin{table}
\caption{Observations during the 2006 superoutburst.}
\label{obs4}  
\begin{center}
\begin{tabular}{lllc}
\hline\hline
Start date$^{*}$ & End date$^{*}$ &$N^{\dagger}$ & Observer$^{\ddagger}$
 \\ 
\hline 

3992.385 & 3992.518 & 187  & M \\
3993.392 & 3993.575 & 207  & M \\
3994.413 & 3994.582 & 208  & M \\
3995.249 & 3995.254 & 10   & O \\
3995.423 & 3995.428 & 8    & M \\
3997.214 & 3997.323 & 502  & N \\
3998.248 & 3998.312 & 119  & O \\
3999.248 & 3999.317 & 124  & O \\
3999.395 & 3999.598 & 246  & M \\
4000.248 & 4000.320 & 185  & N \\ 
4000.291 & 4000.334 & 113  & P \\
4001.238 & 4001.341 & 344  & P \\
4001.420 & 4001.558 & 170  & M \\
4002.224 & 4002.324 & 378  & N \\
4002.233 & 4002.318 & 155  & O \\
4003.234 & 4003.338 & 102  & P \\
4004.369 & 4004.458 & 162  & M \\
4006.205 & 4006.294 & 45   & O \\
4006.230 & 4006.316 & 249  & N \\
4009.228 & 4009.288 & 171  & N \\
4015.252 & 4015.332 & 82   & O \\
4016.217 & 4016.332 & 111  & O \\
4016.225 & 4016.331 & 276  & N \\
4017.234 & 4017.329 & 230  & N \\

\hline
\multicolumn{4}{l}{$^{*}$ HJD-2450000.} \\                           
\multicolumn{4}{l}{$^{\dagger}$ Number of frames.} \\                
\multicolumn{4}{l}{$^{\ddagger}$ See table \ref{obs1} for detail.}\\ 
\end{tabular}
\end{center}
\end{table}

\section{ANALYSIS AND RESULT}
\label{result}

\subsection{\it 2000 Superoutburst}

\begin{figure}
\begin{center}
\FigureFile(80mm,!){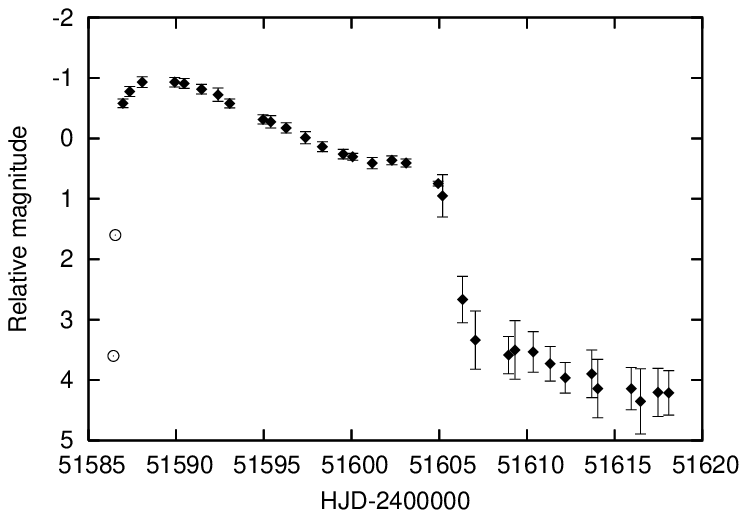}
\end{center}
\caption{Light curve of the 2000 February-March superoutburst. The
 abscissa is HJD, and the ordinate is the relative magnitude to a
 comparison star. The filled diamonds are average magnitudes of the CCD 
observations, and the bars represent the dispersion calculated by using
 data on each day. The open circles represent visual observations.}
\label{00lc}
\end{figure}

The light curve of the 2000 February-March superoutburst is shown 
in figure \ref{00lc}. This
superoutburst ignited on HJD 2451586. The magnitude reached the maximum 
of 10.4 mag on HJD 2451588, and successively the
object faded almost constantly from
HJD 2451589 to HJD 2451600. After that, the object
rebrightened with an amplitude of $\sim0.1$ mag on HJD 2451602. On HJD
2451605, the star entered the rapid fading phase, and for the following
two weeks, there was no evidence of 
rebrightening outbursts which are often observed in WZ Sge-type dwarf novae and
occasionally in SU UMa-type dwarf novae with a short orbital period.

\begin{figure*}
\begin{center}
\FigureFile(140mm,!){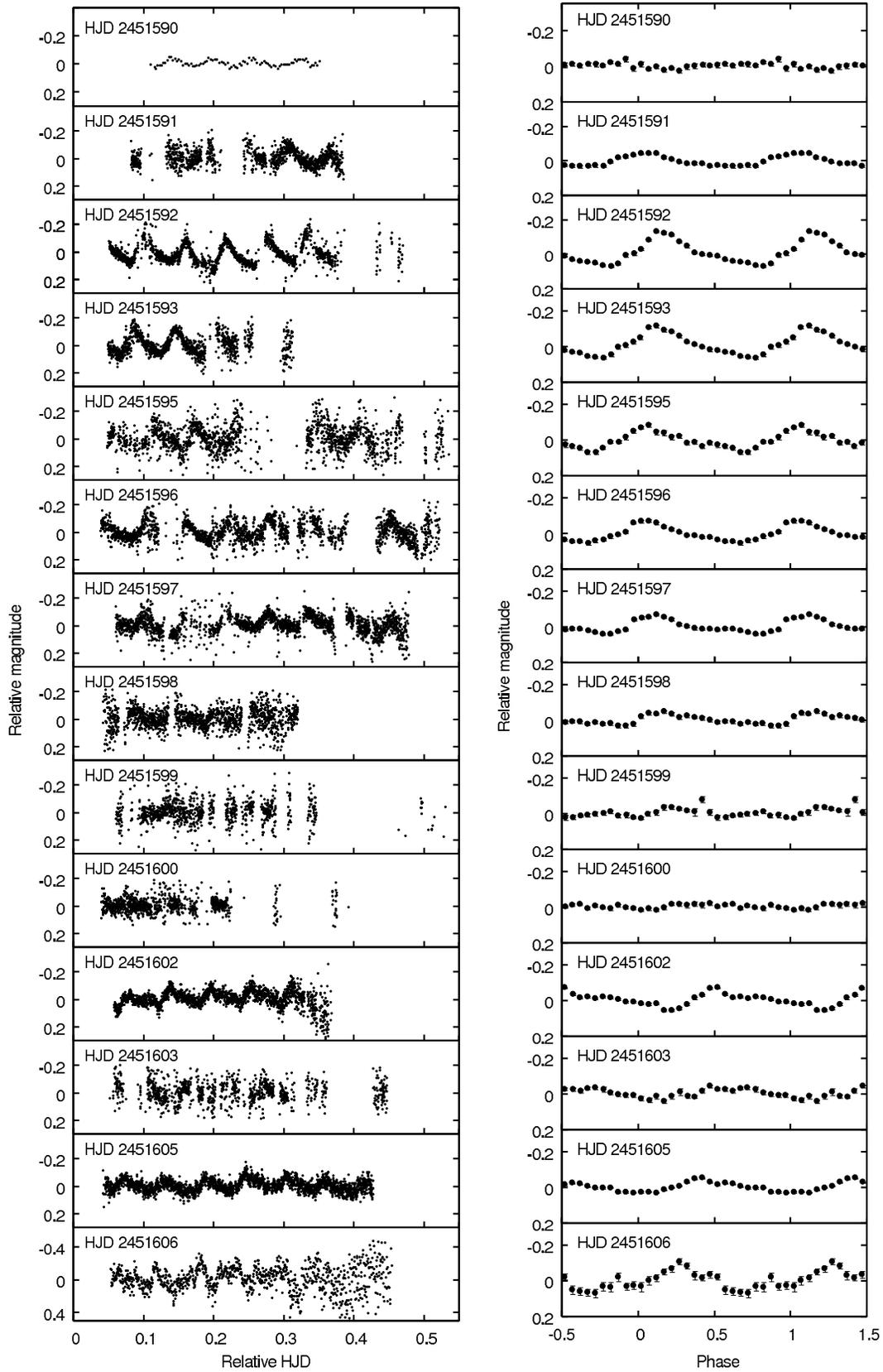}
\end{center}
\caption{Left panel: daily light curves during the 2000 superoutburst.
Right panel: daily phase-averaged light curves of the 2000 
superoutburst folded by 0.058096 days.}
\label{00dailylc}
\end{figure*}

We detected clear ordinary superhumps in separated two instances, from HJD 2451590
to HJD 2451597, and from HJD 2451602 to 2451605 (the left panel of figure
\ref{00dailylc}). The superhump amplitude
reached the first maximum of $\sim0.3$ mag on HJD 2451592, and gradually
declined between HJD 2451593 and 2451600. However, it regrew and reached
the second maximum of $\sim0.2$ mag on HJD 2451602.
On HJD 2451606, at the end of rapid brightness decline of the outburst, 
the humps showed more complex shapes with an amplitude of $\sim0.2$ mag. 

\begin{figure}
\begin{center}
\FigureFile(80mm,!){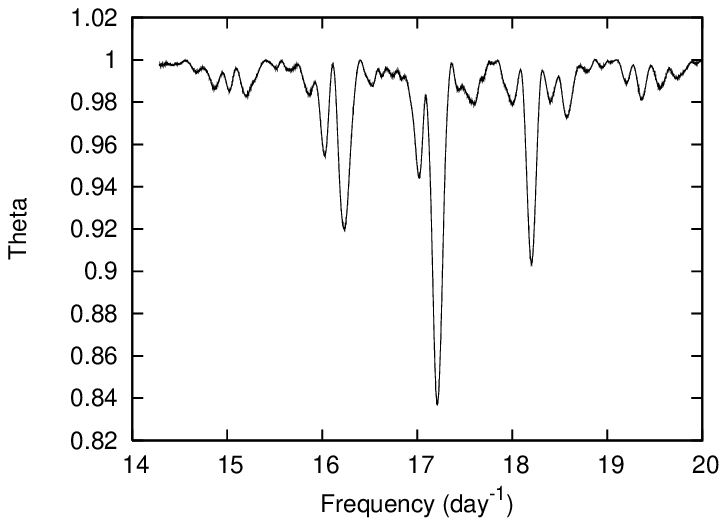}
\end{center}
\caption{Theta-Frequency diagram obtained by the PDM analysis from the
 data of the middle stage of the superhumps during the 2000 superoutburst,
 between HJD 2451592 and 2451600.}
\label{00pdm}
\end{figure}

After subtracting the general trend by fitting first or
second-order polynomials, we carried out the PDM method
(\cite{ste78pdm}) in order to measure the $P_{SH}$ , using the data set between HJD
2451592 and HJD 2451600, from the first maximum of the superhump
amplitude to before the second. From the Theta-Frequency diagram of the
PDM analysis (figure \ref{00pdm}), we determined the
mean superhump frequency to be 17.213(2) day$^{-1}$ ($P_{SH}=0.058096(6)$ days).
The right panel of figure \ref{00dailylc} shows the daily phase-averaged
light curves folded by 0.058096 days, between HJD 2451590 and 2451606. 

We measured the maximum times of the superhumps by eye (table
\ref{00maxtime}). The cycle count ($E$) was set to
be 1 at the first observed superhump maximum. A linear regression yields
a following equation on the maximum timings:

\begin{equation}
\label{00eqc}
HJD_{max}=0.05818(1) \cdot E+2451590.4914(15).
\end{equation} 

\begin{figure}
\begin{center}
\FigureFile(80mm,!){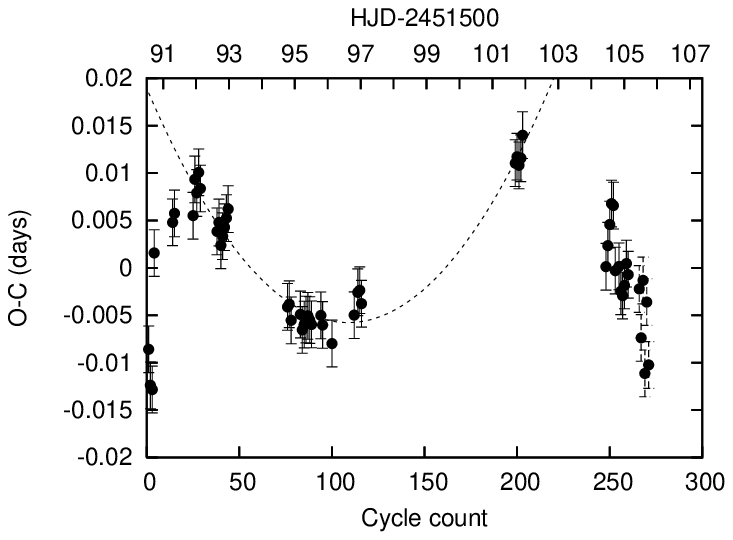}
\end{center}
\caption{$O-C$ diagram of the superhump maximum timings of the 2000
 superoutburst listed in
 table \ref{00maxtime}.  The curved line is obtained by a quadratic polynomial
 fitting to the $O-C$ (equation (\ref{00eqo-c})).}
\label{00o-c}
\end{figure}

Using this equation, we drew an $O-C$ diagram for the maximum timings of
the superhumps (figure \ref{00o-c}).
The $O-C$ diagram represents a decreasing trend of the $P_{SH}$ around
$E=20$, a gradual increase of the $P_{SH}$ in the
middle (around $20<E<200$), and a decrease of the $P_{SH}$ again at the end
(around $E=200$). The $O-C$ diagram between $25<E<203$ can be fitted by
the following quadratic,

\begin{eqnarray}
\label{00eqo-c}
O-C &=& 2.09(9) \times 10^{-6} \cdot E^{2} - 4.54(21) \times 10^{-4} \cdot E 
\nonumber \\
     && +1.89(10) \times 10^{-2}.
\end{eqnarray}

From this equation, the mean $P_{SH}$ derivative in the middle stage
($25<E<203$) is estimated to be $P_{dot}=\dot{P}_{SH}/P_{SH}=7.1(3) \times
10^{-5}$. 

\begin{figure}
\begin{center}
\FigureFile(80mm,!){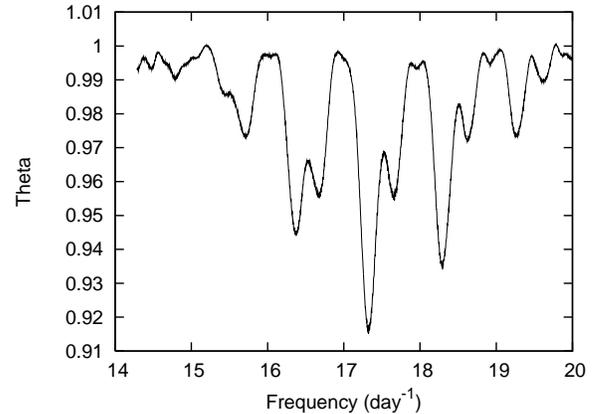}
\end{center}
\caption{Theta-Frequency diagram obtained by the PDM analysis from the
 data of the end stage of the 2000 superoutburst, between HJD 2451603 and 2451606.}
\label{00latepdm1}
\end{figure}

\begin{figure}
\begin{center}
\FigureFile(80mm,!){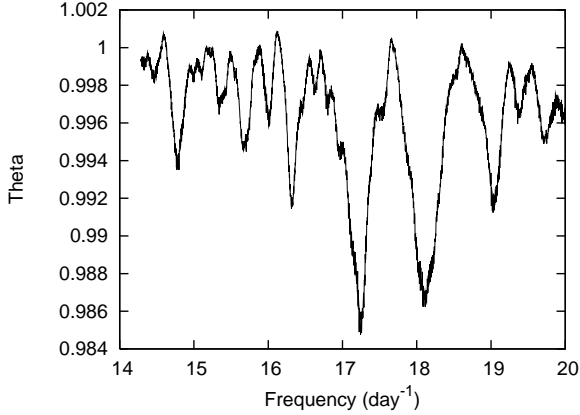}
\end{center}
\caption{Theta-Frequency diagram obtained by the PDM analysis from the
 data after the 2000 superoutburst, between HJD 2451607 and 2451610.}
\label{00latepdm2}
\end{figure}

\begin{figure*}
\begin{center}
\FigureFile(140mm,!){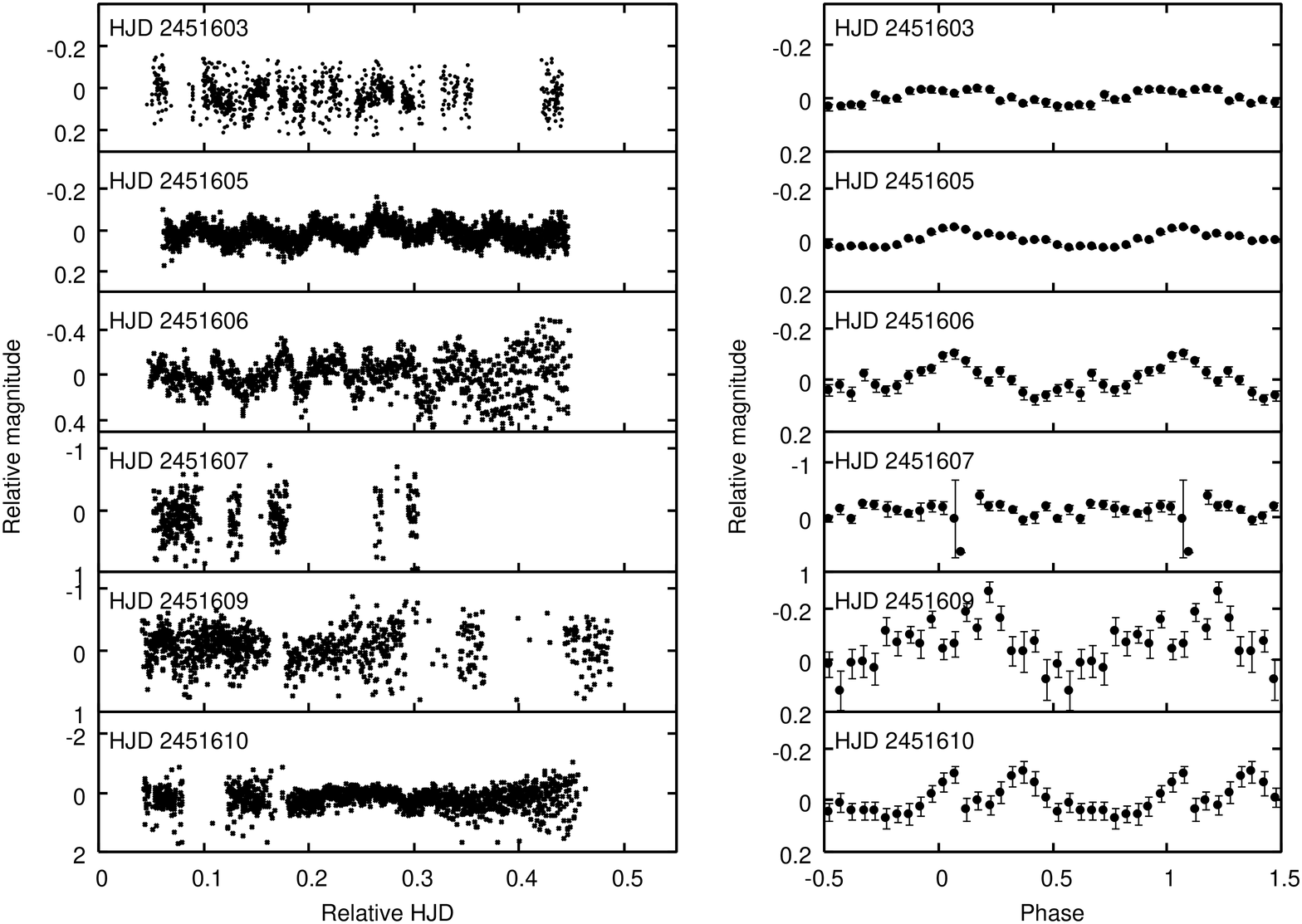}
\end{center}
\caption{Left panel: daily light curves at and after the end stage of the 2000
 superoutburst.
Right panel: daily phase-averaged light curves folded by 0.05771 days.}
\label{00latedailylc}
\end{figure*}

We applied the PDM method to the data sets at the end of this
superoutburst, from HJD 2451603 to 2451606, and from HJD 2451607 to 2451610.
The best estimated frequencies
are 17.327(8) day$^{-1}$ for the former term and 17.248(19) day$^{-1}$,
which correspond to periods of 0.05771(3) and 0.05798(6) days,
respectively (figure \ref{00latepdm1}, \ref{00latepdm2}). 
Figure \ref{00latedailylc} shows the daily
light curves and daily phase-averaged light curves of the end stage of
this outburst (from HJD 2451603 to 2451610). The time of phase zero is
the same as in figure \ref{00dailylc}, and the period used in folding is
0.0057714 days.

\begin{table*}
\caption{Timings of the superhump maxima of the 2000 superoutburst.}
\label{00maxtime}
\begin{center}
\begin{tabular}{lcclcc}  
\hline \hline
 $E^{*}$ & HJD-2400000 & $O-C^{\dagger}$ (days) & $E^{*}$ & HJD-2400000 & $O-C^{\dagger}$ (days) \\ 
\hline 

1  & 51590.54104 & $-$0.0078  & 95  & 51596.01289 & $-$0.0064 \\    
2  & 51590.59543 & $-$0.0116  & 100 & 51596.30185 & $-$0.0084 \\   
3  & 51590.65317 & $-$0.0120  & 112 & 51597.00305 & $-$0.0056 \\   
4  & 51590.72575 &  0.0023  & 114 & 51597.12182 & $-$0.0032 \\  
14 & 51591.31082 &  0.0054  & 115 & 51597.18023 & $-$0.0030 \\  
15 & 51591.36996 &  0.0063  & 116 & 51597.23699 & $-$0.0044 \\  
25 & 51591.95155 &  0.0059  & 199 & 51602.08108 &  0.0093 \\   
26 & 51592.01357 &  0.0098  & 200 & 51602.13994 &  0.0099 \\   
27 & 51592.07032 &  0.0083  & 201 & 51602.19721 &  0.0090 \\   
28 & 51592.13068 &  0.0105  & 202 & 51602.25615 &  0.0098 \\   
29 & 51592.18715 &  0.0088  & 203 & 51602.31676 &  0.0122 \\   
38 & 51592.70628 &  0.0041  & 248 & 51604.92115 & $-$0.0022 \\   
39 & 51592.76541 &  0.0051  & 249 & 51604.98156 & $-$0.0000 \\   
40 & 51592.82119 &  0.0026  & 250 & 51605.04197 &  0.0021 \\    
41 & 51592.88033 &  0.0036  & 251 & 51605.10237 &  0.0043 \\    
42 & 51592.93946 &  0.0045  & 252 & 51605.16034	&  0.0041 \\  
43 & 51592.99860 &  0.0055  & 253 & 51605.21166 & $-$0.0027 \\
44 & 51593.05774 &  0.0064  & 255 & 51605.32847	& $-$0.0022 \\
76 & 51594.90930 & $-$0.0042  & 256 & 51605.38403 & $-$0.0049 \\
77 & 51594.96777 & $-$0.0040  & 257 & 51605.44176 & $-$0.0053 \\             
78 & 51595.02425 & $-$0.0057  & 258 & 51605.50102 & $-$0.0043 \\     
83 & 51595.31579 & $-$0.0051  & 259 & 51605.56151 & $-$0.0020 \\     
84 & 51595.37235 & $-$0.0068  & 260 & 51605.61852 & $-$0.0032 \\     
85 & 51595.43109 & $-$0.0062  & 266 & 51605.96612 & $-$0.0022 \\     
86 & 51595.48982 & $-$0.0057  & 237 & 51606.01914 & $-$0.0074 \\  
87 & 51595.54838 & $-$0.0053  & 268 & 51606.08339 & $-$0.0013 \\  
88 & 51595.60611 & $-$0.0058  & 269 & 51606.13716 & $-$0.0111 \\
89 & 51595.66384 & $-$0.0062  & 270 & 51606.19748 & $-$0.0036 \\
94 & 51595.95572 & $-$0.0054  & 271 & 51606.24902 & $-$0.0102 \\

\hline
\multicolumn{6}{l}{$^{*}$ Cycle count.} \\
\multicolumn{6}{l}{$^{\dagger}$ Using equation (\ref{00eqc}).} \\
\end{tabular}
\end{center}
\end{table*}

\subsection{\it 2002 Superoutburst}

\begin{figure}
\begin{center}
\FigureFile(80mm,!){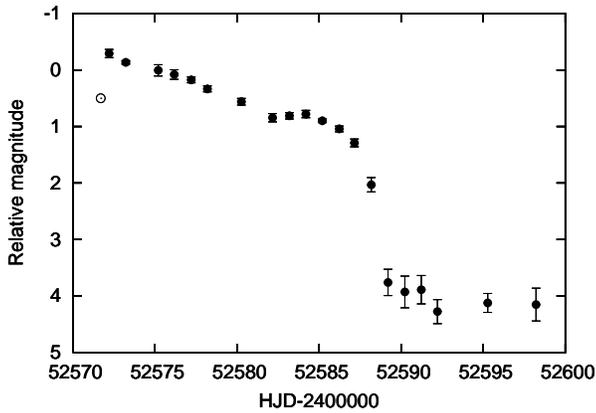}
\end{center}
\caption{Light curve of the 2002 October-November superoutburst. The
 abscissa is HJD, and the ordinate is the relative magnitude to a
 comparison star. The filled diamonds are average magnitudes of the CCD 
observations and the bars represent the dispersion calculated by using
 data on each day. The open circle represents a visual observation.}
\label{02lc}
\end{figure}

Figure \ref{02lc} shows the light curve of the 2002 October-November
superoutburst. This superoutburst was detected at 11.3 mag on HJD
2452571, and the magnitude reached the maximum of
10.5 mag on HJD 2452572. The object faded almost constantly
from HJD 2452575 to
HJD 2452580. After the constant decline, the object
rebrightened with an amplitude of $\sim0.1$ mag on HJD 2452584. On HJD
2452589, this object came back to quiescence, and no rebrightening outburst was
observed for following the few weeks. 

\begin{figure*}
\begin{center}
\FigureFile(130mm,!){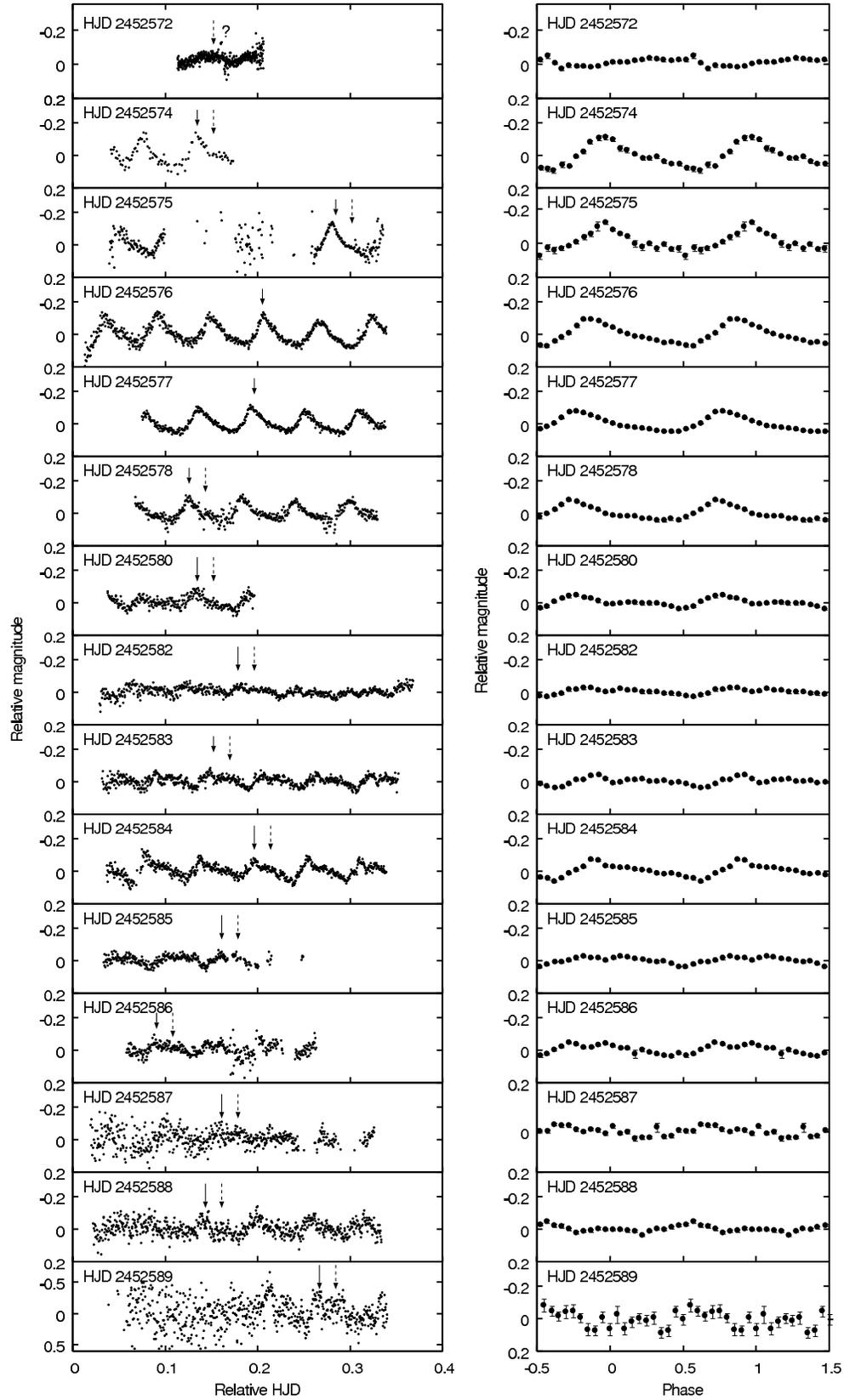}
\end{center}
\caption{Left panel: daily light curves during the 2002 superoutburst. The solid
arrows and the dashed arrows indicate positions of two kinds of humps, 
respectively (see section \ref{discussion3} in detail).
Right panel: daily phase-averaged light curves folded by 0.058278 days.}
\label{02dailylc}
\end{figure*}

Faint modulations with an amplitude of $\sim$0.05 mag which are probably
the seed of the superhump can be seen in the light curve on HJD 2452572,
and clear superhumps were observed from HJD 2452574
to 2452588 (the left panel of figure \ref{02dailylc}). The superhump
amplitude reached
$\sim$0.2 mag on HJD 2452574, and gradually
decreased after that (between HJD 2452575 and 2452582). It, then, grew
up again around HJD 2452584.
There are obscure humps with an amplitude of $\sim$0.5 mag in the light
curve of HJD 2452589, near the quiescence level.

\begin{figure}
\begin{center}
\FigureFile(80mm,!){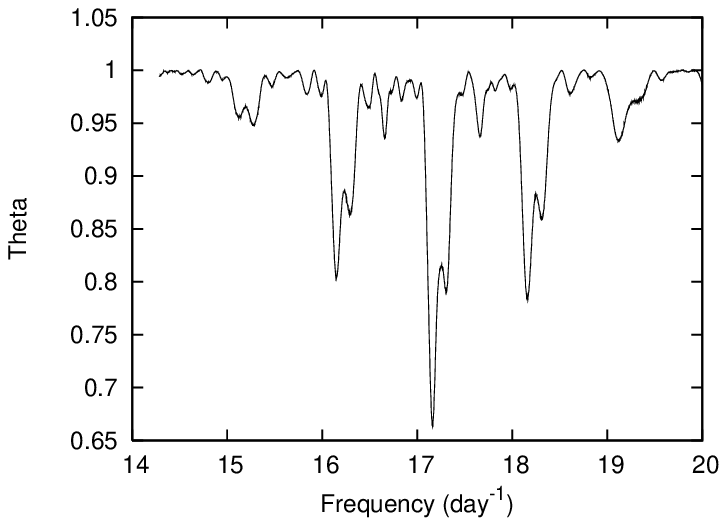}
\end{center}
\caption{Theta-Frequency diagram obtained by the PDM analysis from the
 data of the middle stage of the superhumps during the 2002
 superoutburst, between HJD 2452574 and 2452582.}
\label{02pdm}
\end{figure}

We determined the average superhump frequency to be 17.161(2) day$^{-1}$
($P_{SH}=0.058271(5)$ days) by applying the PDM method to data between
HJD 2452574 and 2452582, from the first amplitude maximum to before the
second (figure \ref{02pdm}). 
The right panel of figure \ref{02dailylc} represents the daily
phase-averaged light curves
folded by 0.058271 days from HJD 2452572 to 2452588.

We measured the maximum times of the superhumps by eye (table
\ref{02maxtime}). A linear regression yields the following equation:

\begin{equation}
\label{02eqc}
HJD_{max}=0.05826(1) \cdot E+2452574.3454(15).
\end{equation} 

\begin{figure}
\begin{center}
\FigureFile(80mm,!){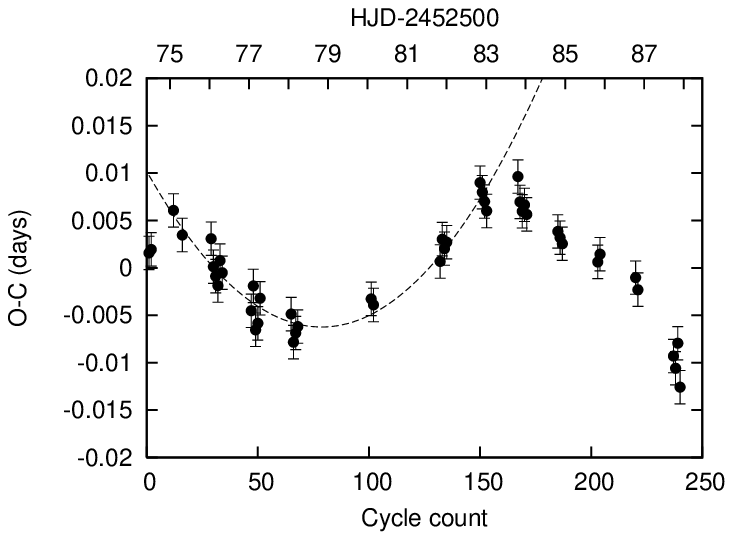}
\end{center}
\caption{$O-C$ diagram of the superhump maximum timings of the 2002
 superoutburst listed in
 table \ref{02maxtime}. The curved line is obtained by a quadratic polynomial
 fitting to the $O-C$ (equation (\ref{02eqo-c})).}
\label{02o-c}
\end{figure}

Using equation \ref{02eqc}, we obtained an $O-C$ diagram for the timings of
the superhump
maxima of this superoutburst (figure \ref{02o-c}).
Figure \ref{02o-c} indicates that the $P_{SH}$ decreased around $E=10$,
and gradually increased around $10<E<150$,
and, subsequently, decreased again around $E=150$. For
$12<E<153$, the best fitting quadratic equation is given by:

\begin{eqnarray}
\label{02eqo-c}
O-C &=& 2.65(18) \times 10^{-6} \cdot E^{2} - 4.18(33) \times 10^{-4} \cdot E 
\nonumber \\
     && +1.02(12) \times 10^{-2}.
\end{eqnarray}

From this equation, the mean $P_{SH}$ derivative for $12<E<153$ is
estimated to be $P_{dot}=9.1(6) \times 10^{-5}$. 

\begin{figure}
\begin{center}
\FigureFile(80mm,!){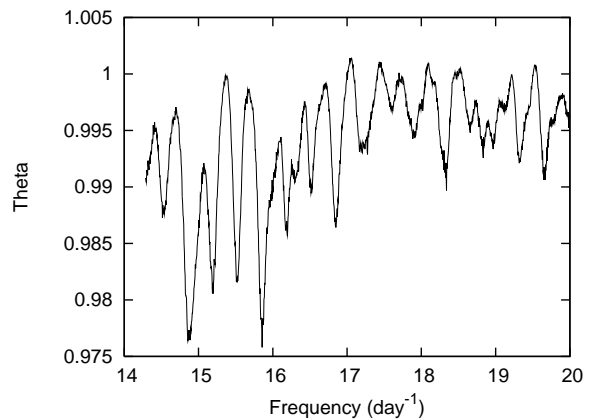}
\end{center}
\caption{Theta-Frequency diagram obtained by the PDM analysis from the
 data after the plateau phase of the 2002 superoutburst between HJD
 2452589 and 2452592.}
\label{02latepdm}
\end{figure}

Using the data set after the rapid decline phase (HJD 2452589 and 2452592),
we carried out the PDM analysis. No plausible frequency was
detected, however (figure \ref{02latepdm}).

\begin{table*}
\caption{Timings of the superhump maxima of the 2002 superoutburst.}
\label{02maxtime}
\begin{center}
\begin{tabular}{lcclcc}  
\hline \hline
 $E^{*}$ & HJD-2400000 & $O-C^{\dagger}$ (days) & $E^{*}$ & HJD-2400000 & $O-C^{\dagger}$ (days) \\ 
\hline 

1   & 52574.40521 & 0.0041   & 134 & 52582.15419 & 0.0021 \\
2   & 52574.46385 & 0.0044   & 135 & 52582.21313 & 0.0028 \\ 
12  & 52575.05056 & 0.0064   & 150 & 52583.09330 & 0.0090 \\  
16  & 52575.28101 & 0.0038   & 151 & 52583.15056 & 0.0080 \\  
29  & 52576.03799 & 0.0033   & 152 & 52583.20782 & 0.0070 \\  
30  & 52576.09330 & 0.0004   & 153 & 52583.26508 & 0.0060 \\  
31  & 52576.15056 & $-$0.0006  & 167 & 52584.08436 & 0.0096 \\  
32  & 52576.20782 & $-$0.0016  & 168 & 52584.13994 & 0.0070 \\  
33  & 52576.26871 & 0.0010   & 169 & 52584.19721 & 0.0060 \\  
34  & 52576.32570 & $-$0.0003  & 170 & 52584.25614 & 0.0066 \\  
47  & 52577.07905 & $-$0.0043  & 171 & 52584.31341 & 0.0056 \\  
48  & 52577.13994 & $-$0.0017  & 185 & 52585.12723 & 0.0038 \\  
49  & 52577.19357 & $-$0.0063  & 186 & 52585.18485 & 0.0032 \\   
50  & 52577.25251 & $-$0.0056  & 187 & 52585.24246 & 0.0025 \\   
51  & 52577.31341 & $-$0.0030  & 203 & 52586.17270 & 0.0006 \\   
65  & 52578.12737 & $-$0.0047  & 204 & 52586.23177 & 0.0014 \\   
66  & 52578.18268 & $-$0.0076  & 220 & 52587.16145 & $-$0.0011 \\
67  & 52578.24190 & $-$0.0067  & 221 & 52587.21844 & $-$0.0024 \\
68  & 52578.30084 & $-$0.0060  & 237 & 52588.14358 & $-$0.0094 \\  
101 & 52580.22633 & $-$0.0031  & 238 & 52588.20056 & $-$0.0107 \\  
102 & 52580.28394 & $-$0.0038  & 239 & 52588.26145 & $-$0.0081 \\
132 & 52582.03631 & 0.0007   & 240 & 52588.31508 & $-$0.0127 \\  
133 & 52582.09693 & 0.0031 &&& \\

\hline
\multicolumn{6}{l}{$^{*}$ Cycle count.} \\
\multicolumn{6}{l}{$^{\dagger}$ Using equation (\ref{02eqc}).} \\
\end{tabular}
\end{center}
\end{table*}

\subsection{\it 2006 Superoutburst}

\begin{figure}
\begin{center}
\FigureFile(80mm,!){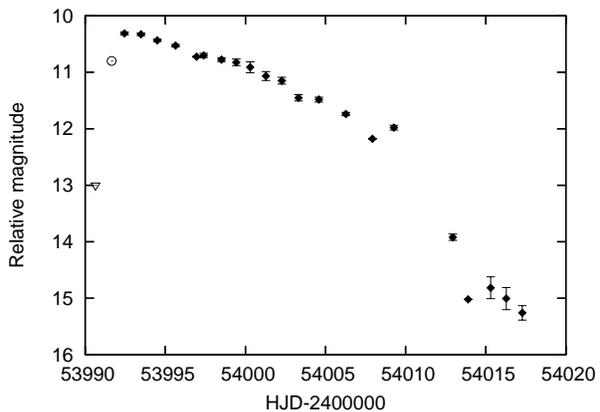}
\end{center}
\caption{Light curve of the 2006 September superoutburst. The
 abscissa is HJD, and the ordinate is the relative magnitude to a
 comparison star. The filled diamonds are average magnitudes of the CCD 
observations and the bars represent the dispersion calculated by using
 data on each day. The open circles and the bottom triangle mean the
 visual and the upper limit, respectively.}
\label{06lc}
\end{figure}

The light curve of the 2006 September superoutburst
is shown in figure \ref{06lc}. This
superoutburst started on HJD 2453991, and the magnitude became a maximum
of 10.3 mag on HJD 245392. This object also showed a rebrightening at the late
stage of this outburst (around HJD
2454009) as in the 2000 and the 2002 one.

\begin{figure*}
\begin{center}
\FigureFile(150mm,!){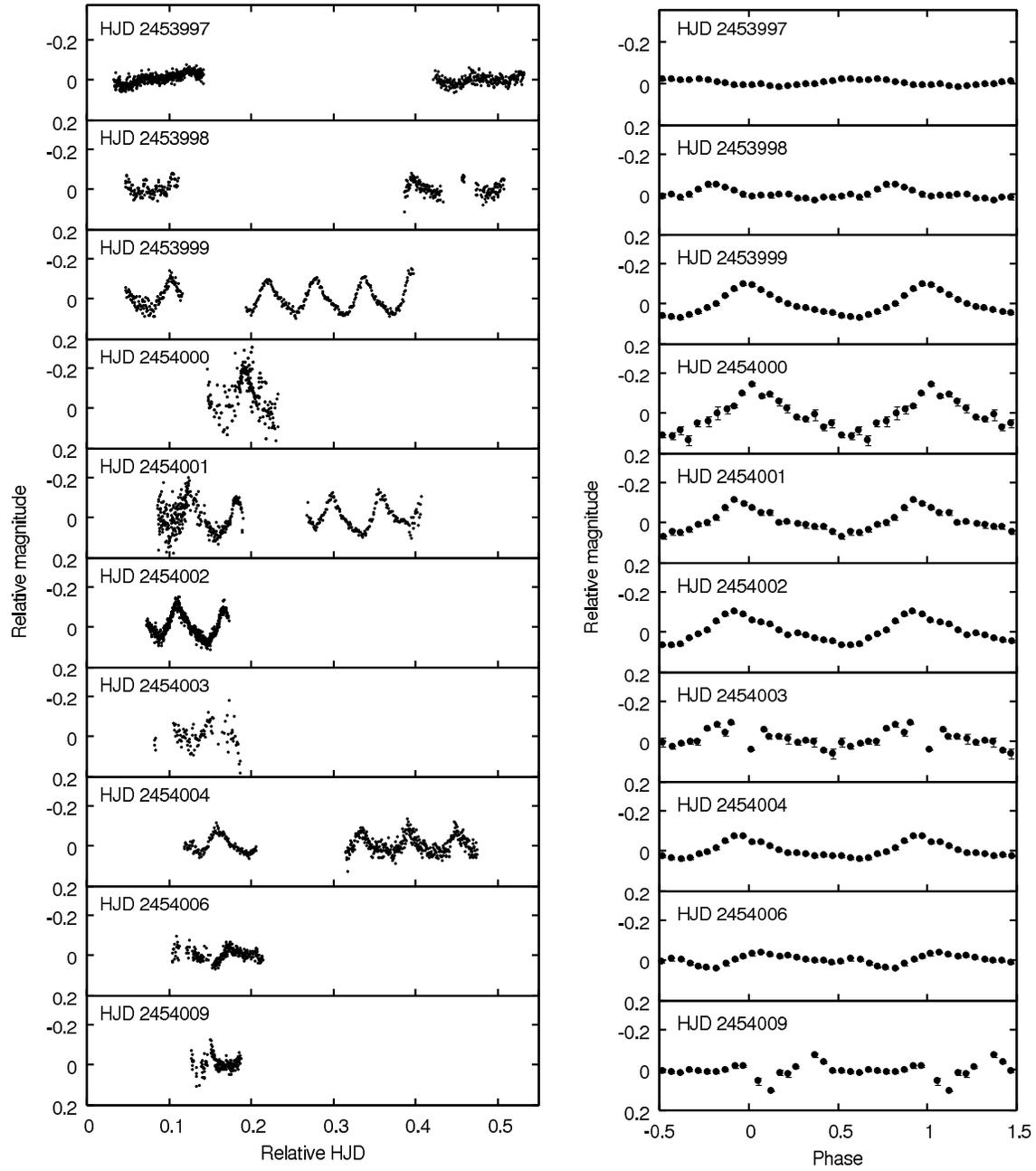}
\end{center}
\caption{Left panel: daily light curves during the 2006 superoutburst.
Right panel: daily phase-averaged light curves folded by 0.058063 days.}
\label{06dailylc}
\end{figure*}

\begin{figure}
\begin{center}
\FigureFile(80mm,!){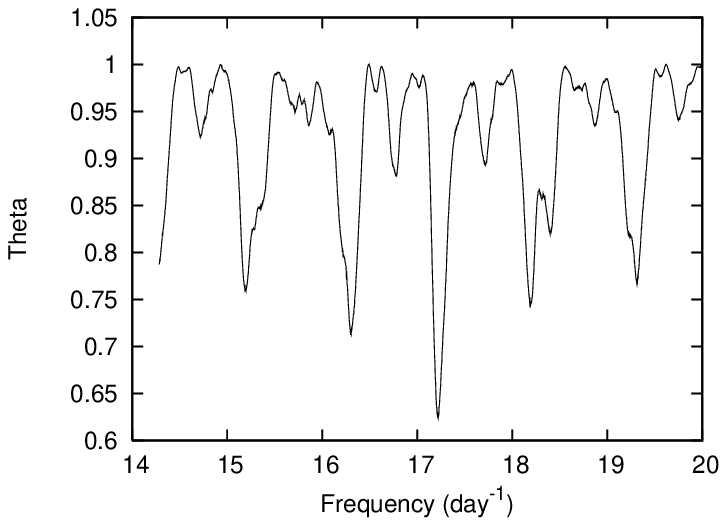}
\end{center}
\caption{Theta-Frequency diagram obtained by the PDM analysis from the
 data of the middle stage of the superhumps during the 2006
 superoutburst, between HJD 2454000 and 2454009.}
\label{06pdm}
\end{figure}

Clear superhumps have been observed between HJD 2453997 and
 2454009 (the left panel of figure \ref{06dailylc}).
The superhump amplitude was increasing until HJD 2454000, and decreasing
thereafter. 
Figure \ref{06pdm} shows the Theta-Frequency diagram of the PDM analysis
of the data between HJD 2454000 and 2454009. After the first amplitude
maximum, we determined the most probable superhump frequency
to be 17.226(3) day$^{-1}$ ($P_{SH}=0.058050(10)$ days). 
The daily phase-averaged light curves
folded by 0.058050 days between HJD 2453997 and 2454009 are exhibited in
the right panel of figure \ref{06dailylc}.

We measured the maximum times of the superhumps by eye
(table \ref{06maxtime}). A linear regression yields a following equation
on the maximum timings:

\begin{equation}
\label{06eqc}
HJD_{max}=0.05816(3) \cdot E+2453997.6109(26).
\end{equation} 

\begin{figure}
\begin{center}
\FigureFile(80mm,!){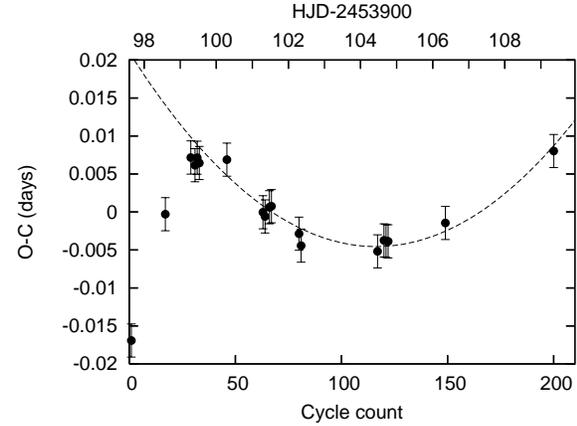}
\end{center}
\caption{$O-C$ diagram of the superhump maximum timings of the 2006
 superoutburst listed in
 table \ref{06maxtime}. The curved line is obtained by a quadratic polynomial
 fitting to the $O-C$ (equation (\ref{06eqo-c})).}
\label{06o-c}
\end{figure}

We calculated the $O-C$ values of the maximum timings of the superhumps,
based on equation \ref{06eqc}, and plotted them on figure \ref{06o-c}.
This figure shows a decreasing trend of the $P_{SH}$ around $E=30$, and
a gradual increase around $E>40$. The $O-C$ diagram between
$46<E<200$ can be fitted by the following quadratic,

\begin{eqnarray}
\label{06eqo-c}
O-C &=& 1.89(18) \times 10^{-6} \cdot E^{2} - 4.40(42) \times 10^{-4} \cdot E 
\nonumber \\
     && +2.10(23) \times 10^{-2}.
\end{eqnarray}

This equation yields the mean $P_{SH}$ derivative for
$46<E<200$, $P_{dot}=6.5(0.6) \times 10^{-5}$.

\begin{table}
\caption{Timings of superhump maxima of the 2006 superoutburst.}
\label{06maxtime}
\begin{center}
\begin{tabular}{lcc}  
\hline \hline
 $E^{*}$ & HJD-2400000 & $O-C^{\dagger}$ (days) \\ 
\hline 

1   & 53997.65210 & $-$0.01642 \\   
17  & 53998.59922 &  0.00044 \\   
29  & 53999.30454 &  0.00806 \\
31  & 53999.41984 &  0.00708 \\
32  & 53999.47898 &  0.00808 \\
33  & 53999.53643 &  0.00739 \\
46  & 54000.29290 &  0.00803 \\
63  & 54001.27460 &  0.00133 \\
64  & 54001.33221 &  0.00080 \\
66  & 54001.44670 &  0.00203 \\  
67  & 54001.50590 &  0.00218 \\
80  & 54002.26043 & $-$0.00124 \\
81  & 54002.31700 & $-$0.00281 \\
117 & 54004.41090 & $-$0.00305 \\ 
120 & 54004.58660 & $-$0.00157 \\ 
121 & 54004.64360 & $-$0.00162 \\ 
122 & 54004.70120 & $-$0.00168 \\ 
149 & 54006.27453 &  0.00112 \\
200 & 54009.24991 &  0.01131 \\

\hline
\multicolumn{3}{l}{$^{*}$ Cycle count.} \\
\multicolumn{3}{l}{$^{\dagger}$ Using equation (\ref{06eqc}).} \\
\end{tabular}
\end{center}
\end{table}

\subsection{\it Outbursts of SW UMa}
\label{longterm}

\begin{figure*}
\begin{center}
\FigureFile(160mm,!){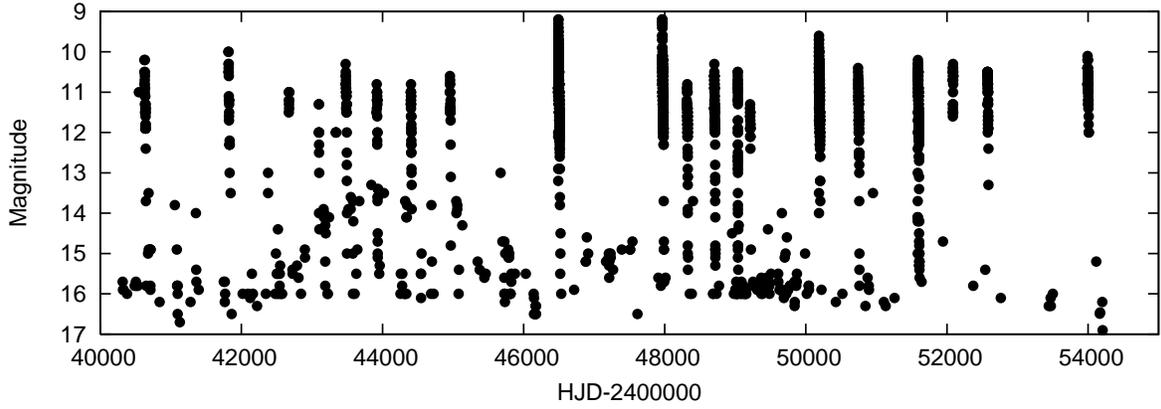}
\end{center}
\caption{Long-term light curve of SW UMa. The abscissa is HJD,
 and the ordinate is magnitude. The filled circles represent visual observations.}
\label{longlc}
\end{figure*}

\begin{figure}
\begin{center}
\FigureFile(80mm,!){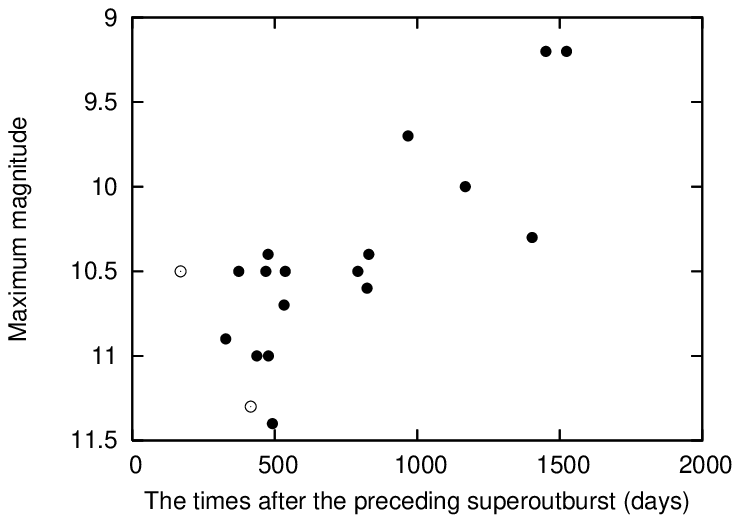}
\end{center}
\caption{The relation between the maximum magnitude of an outburst and
the time after the preceding superoutburst of SW UMa.
The filled circles and the open circles represent  superoutbursts,
and normal outbursts, respectively.}
\label{outburst}
\end{figure}

Figure \ref{longlc} exhibits the long-term light curve of
SW UMa, constructed from enormous number of visual observations reported to
AAVSO in the
past 40 years. We investigated the duration and maximum magnitude of
each outburst using VSNET, AAVSO, and AFOEV archival data
(table \ref{properties}), and checked them against previous works (table
1 in \cite{wen86DNcycle}, table 4 in \cite{how95swumabcumatvcrv}). 
Most outbursts of SW UMa are a superoutburst,
but the 1976 and the 1993 February outbursts are the only two 
certain normal outbursts, judging from their durations.
A single observation of $V=12.2$ on HJD 2443152 was reported to AFOEV,
which seems an outburst, but can not be classified because of the lack
of observations.
Figure \ref{outburst} shows the relation between the maximum magnitude
of an outburst and the time after the preceding superoutburst.

\begin{table*}
\caption{Properties of outbursts.}
\label{properties}
\begin{center}
\begin{tabular}{lccccc}  
\hline \hline
Year & Start Date$^{*}$ & End Date$^{*}$ & Duration (days)$^{\dagger}$ &
 $m_{max}^{\dagger}$  \\ 
\hline 

1970 & 40624 & 40645 & 22 & 10.2  \\
1973 & 41814-41816 & 41834 & 19-21 & 10.0  \\
1975 & 42662-42663 & 42679-42682 & 17-21 & 10.6  \\
1976$^{\S}$ & 43098 & 43100 & 3 & 11.3  \\
1977 Jan. & \multicolumn{3}{c}{HJD 2443152 (single obs.)} & 12.2 \\ 
1977 Dec. & 43474-43478 & 43494 & 17-21 & 10.5  \\
1979 & 43917 & 43930 & 14 & 10.8  \\
1980 & 44396-44400 & 44412 & 13-17 & 11.0  \\
1981 & 44945-44953 & 44967 & 15-23 & 10.7  \\
1986 & 46491 & 46513 & 23 & 9.2  \\
1990 & 47964 & 47985 & 22 & 9.2  \\
1991 & 48313 & 48327 & 15 & 10.9  \\
1992 & 48701 & 48716 & 17 & 10.5  \\
1993 Feb.$^{\S}$ & 49035 & 49038 & 4 & 10.5  \\
1993 Aug. & 49208 & 49217 & 10 & 11.4  \\
1996 & 50185 & 50203 & 19 & 9.7  \\
1997 & 50740 & 50756 & 17 & 10.5  \\
2000 & 51586 & 51605 & 20 & 10.4  \\
2001 & 52082 & 52090-52098 & 9-17 & 10.4  \\
2002 & 52571 & 52588 & 18 & 10.5  \\
2006 & 53991 & 54012 & 23 & 10.3  \\

\hline
\multicolumn{5}{l}{$^{*}$ HJD-2400000.} \\
\multicolumn{5}{l}{$^{\dagger}$ Days of $V \ge 13$.} \\
\multicolumn{5}{l}{$^{\ddagger}$ Maximum magnitude of the outburst.} \\
\multicolumn{5}{l}{$^{\S}$ Normal outburst.} \\
\end{tabular}
\end{center}
\end{table*}

\subsection{QPOs}

\begin{figure*}
\begin{center}
\FigureFile(130mm,!){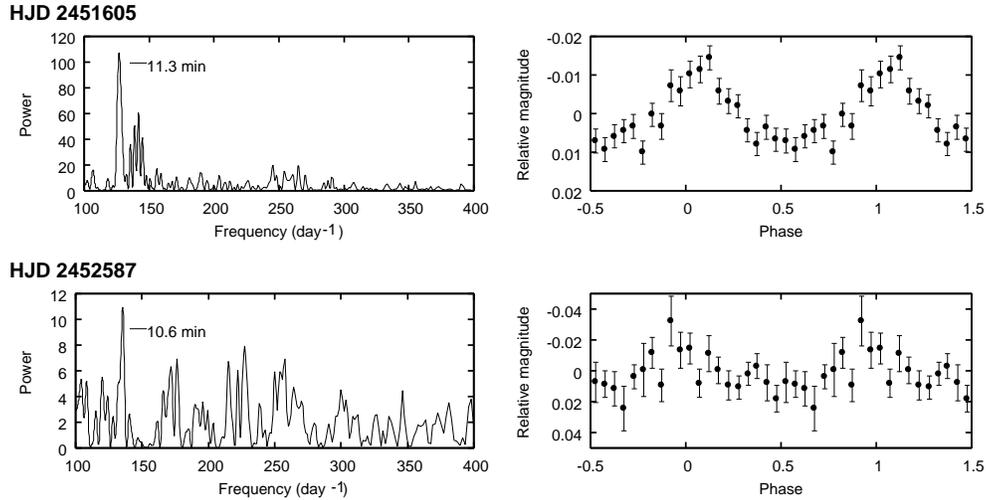}
\end{center}
\caption{Results of QPO analyses. The left and right panels are
 the power spectra and the phase-averaged light curves for each day at
 the end stage of the 2000, and the 2002 superoutbursts.
The best estimated periods used in foldings are
11.3 min and 10.6 min in HJD 2451605 and 2452587, respectively.}
\label{qpo}
\end{figure*}

After subtracting the mean superhump profile, we constructed power
spectra of each day during the 2000, the 2002, and the 2006
superoutbursts in order to detect QPOs. 
The left panels in figure \ref{qpo} are the power spectra from the data
of HJD 2451605 and 2452587, and the strongest signals correspond to 11.3 min
and 10.6 min, respectively. The right panels in figure \ref{qpo}
represent the phase-averaged light curves folded by the best estimated period.

\section{DISCUSSION}
\label{discussion}

\subsection{\it Light Curve Properties}
\label{discussion1}

During three superoutbursts we investigated, the brightness decline
was slowed down when the superhumps developed (see figures, \ref{00lc},
\ref{00dailylc}, \ref{02lc}, \ref{02dailylc}, \ref{06lc}, and
\ref{06dailylc}). Especially, there was a
slight rebrightening at the same time when the
regrowth of the superhump amplitude occurred during the 2000 and the
2002 superoutbursts. Since the emitted
energy during an outburst is essentially given by the amount of mass
accreted inward, this phenomenon indicates the regrowth of disk
eccentricity and the resurgence of inward flow of material caused by the
tidal torque in the accretion disk at the end stage of these outbursts.
In the case of the 2006 superoutburst, although a rebrightening was
observed at the end end stage of the outburst (HJD 2454009), we can
trace the variation of the superhump amplitude due to the lack of the
observation (figure \ref{06lc}). A regrowth of the superhump would have
occurred during this superoutburst as the previous two superoutbursts.

Some SU UMa-type dwarf novae, e.g. SW UMa
\citep{how95swumabcumatvcrv} and WX Cet (\cite{kat95wxcet},
\cite{ste07wxcet}), show more various scales of superoutbursts than
ordinary SU UMa-type dwarf novae. 
\citet{osa95wzsge} indicated that if $\dot{M}$ is considered to be
constant, the amplitude of a superoutburst
is proportional to $t_{wait-S}$, where $\dot{M}$ is the
mass transfer rate from the secondary, and $t_{wait-S}$ is the waiting
time for the next superoutburst after which any normal outburst can trigger a
superoutburst. Figure \ref{outburst} shows that the maximum magnitude of
a superoutburst has roughly linear correlation with the time after the 
preceding superoutburst (recurrence time, $t_{S}$), which seems to
confirm the prediction by \citet{osa95wzsge}, although $t_{wait-S}$ is not
strictly equal to $t_{S}$. On the other hand, superoutbursts after
almost the same recurrence time showed somewhat different maximum magnitudes. 
The mass transfer rate can not be always regarded as constant.

\citet{osa95wzsge} also provided a model for WZ Sge-type dwarf novae which
are a very inactive subgroup of SU UMa stars, and thought to be an end
product of the CV evolution. Their model explained some of the observational
properties of WZ Sge stars; for example, the extremely long recurrence time of
superoutbursts (1000-10000 days), no (or few) normal outbursts, and the
large amplitude of superoutbursts (6-8 mag).
According to \citet{osa95wzsge}, for dwarf novae with a moderately
low mass transfer rate, the recurrence time of the normal
outbursts, $t_{N}$, is proportional to the inverse square of $\dot{M}$,
and $t_{wait-S}$ is proportional to the inverse
of $\dot{M}$. They presented the number of outbursts in one supercycle as 

\begin{equation}
\label{number}
$the number of normal outbursts$ \propto t_{wait-S}/t_{N} \propto \dot{M},
\end{equation}

\noindent provided that $t_{wait-S} > t_{N}$.
They suggested that if the mass transfer rate further decreases, we have
$t_{wait-S} < t_{N}$. Then every normal outburst
triggers a superoutburst; this case corresponds to
WZ Sge stars.
In SW UMa, the recurrence times of superoutbursts are about 400 to 1500
days, and these are
somewhat shorter compared with WZ Sge stars (for example, about 30 years
for WZ Sge itself). However, there is no (or few) normal outbursts in SW
UMa, like WZ Sge stars. Therefore, this object
has a relatively small mass transfer rate corresponding to
$t_{wait-S} \sim t_{N}$, and it is going to evolve into WZ Sge type.
The short orbital period of SW UMa is also consistent with the
conclusion. 
This is also applicable in some SU UMa-type dwarf novae, e.g. WX Cet
\citep{ste07wxcet} and BC UMa \citep{mae07bcuma}.

\subsection{\it Superhump Evolution}

We will compare the superhump evolution during well observed
superoutbursts of SW UMa, the 2000, the 2002, the 2006,
and the 1996 superoutbursts. The delay time of appearance of the
superhumps after the onset of the
superoutburst was somewhat different in these four superoutbursts: the delay
was about 4 days for the 2000 one (maximum magnitue, $m_{max}=10.4$), 2
days for the 2002 one ($m_{max}=10.5$), 6 days for
the 2006 one ($m_{max}=10.3$), and 4 days for the 1996 one
($m_{max}=9.7$, \cite{nog98swuma}). \citet{lub91SHa} showed that
the growth rate of the 3:1 resonance is proportional to $q^{2}$
(mass ratio $q=M_{2}/M_{1}$, where $M_{1}$ and $M_{2}$ are the masses of the
primary white dwarf and the secondary star, respectively). This
can not explain the variation of the delay time in the same object,
however. \citet{osa03DNoutburst}, on the other hand, proposed that the
suppression of the 3:1 resonance by the 2:1 resonance is the main
cause of the long delay time in superoutbursts of WZ Sge stars.
As mentioned by \citet{kat08wzsgelateSH}, this interpretation predicts the
variation of the delay time in the same object with a quite
low mass ratio depending on the maximum radius of the accretion disk at
the onset of superoutbursts. They also
suggested that the suppresion by the 2:1 resonance seems the cause of
the variation of the
delay time in SW UMa. The superoutbursts of SW UMa we investigated
seems to agree with the trend suggested by \citet{kat08wzsgelateSH}. 
Our investigation also indicates that 
the delay time was different, but the superhumps showed the same
evolutional trend after
their appearance during these four superoutbursts as below.

During the 2000 superoutburst, after the appearance of the superhumps,
the superhump amplitude reached the first maximum on HJD 2451592,
gradually declined between
HJD 2451593 and 2451600, then started to regrow, and finally reached 
the second maximum on HJD 2451602 (the left panel of figure \ref{00dailylc}). 
The $O-C$ diagram on the maximum timings of the superhumps represents
that the superhump period decreased around HJD 2451592, increased with
the $P_{dot}=7.1(3) \times 10^{-5}$ 
between HJD 2451592 and 2452602, and decreased again around HJD 2451602
(figure \ref{00o-c}).

In the case of the 2002 superoutburst, although the first developing
phase of the superhump was not observed well, the superhumps
traced the same evolution as those during the 2000 one.
The superhump period decreased around HJD 2452575 (probably the
superhump amplitude reached the first maximum around here), and then increased with
the $P_{dot}=9.1(6) \times 10^{-5}$ from the first maximum until
the second maximum of the superhump amplitude (between HJD 2452575 and 2452584)
. Successively, it decreased again around HJD 2452584 (figure
\ref{02dailylc} and \ref{02o-c}).

The second maximum of the superhump amplitude accompanying the decrease
of the superhump period at the end of the bright plateau stage could
not be found during the 2006 superoutburst, due to the lack of the
observation (see also section \ref{discussion1}).  At the early
and middle stage, however, the superhumps also showed the same
trend as those during the 2000 superoutburst.
The superhump period decreased around the time of the amplitude maximum,
and it gradually
increased thereafter with the $P_{dot}=6.5(6) \times 10^{-5}$
(figure \ref{06dailylc} and \ref{06o-c}).

Further, there is a sign of the same evolution of the superhumps
during the 1996 one, although it was observed only at the middle stage. 
The superhump period increased with the $P_{dot}=8.9(1.0)$ days
cycle$^{-1}$ while the superhump amplitude were decreasing, and
the amplitude regrew at the end of the increase of the period
(see figure 2 and 3 in \cite{sem97swuma}).

After their appearance, the superhumps showed the same evolutional trend
during the 2000 and the 2002
superoutbursts. This evolution agrees with the phenomenological
suggestion by \citet{soe09asas1600}:the increase of the superhump period
accompanies the regrowth of the amplitude, and during a
superoutburst with a regrowth of the superhump amplitude, the superhump
period decreases around the first maximum of the
amplitude, successively gradually increases until the
second maximum of the amplitude, and decreases again after that.
Since the superhumps during
the 2006 and the 1996 ones showed the increase of the period
and some similarities with those during the 2000 and the 2002 ones,
the suggestion on the superhump evolution by \citet{soe09asas1600} also seems
applicable to the superhumps during these two superoutbursts.
This same trend of superhump evolution during these four
superoutbursts indicates that superhump evolution may be governed by
invariable binary parameters, such as the mass ratio, the orbital
period, and so on.

However, if the superhump evolution is different in superoutbrursts of
the same object, it is suggested to be affected by the variable
features, e.g., the recurrence time of the outburst
\footnote{Although \citet{uem05tvcrv} found a difference of the
derivative of the superhump period between during the 2001 and the 2004
superoutbursts of TV Crv, Kato et al. (in prep.) reanalysed the
data, and suggest that the interpretation
by \citet{uem05tvcrv} could not be confirmed.}.
As described above, 
the superhumps showed an increase of their period during all well
observed superoutbursts of SW UMa.
\citet{kat98super} suggested that the outward propagation of the
eccentricity generated at the 3:1 resonance radius leads to the increase
of the superhump period.
This predicts that, if the accretion disk does not expand
sufficiently beyond the 3:1 resonance, the eccentricity can not
propagate outward, and the superhump period can not increase.
Since the superhumps showed a period increase, the accretion
disk seemed to expand enough in the superoutbursts of SW UMa we
investigated. These superoutbursts took place more than 500 days after
the preceding superoutburst
(table \ref{properties} and figure \ref{outburst}).
It may be possible, however, that superoutbursts occur with a
minimum recurrence time of $\sim400$ days before sufficient mass
is accumulated in the accretion disk. 
If different superhump evolutions are observed in different 
superoutbursts of one object, this will be a key to reveal which binary
parameters govern the superhump evolution. 
Therefore, observation are required on superoutbursts of
SW UMa which take place after shorter recurrence time in the future.

Detailed analyses have been also carried out in a few other SU
UMa-type dwarf novae repeatedly during different superoutbursts; e.g.,
the 1989 one \citep{odo91wzsge}, the 1998 one (\cite{kat01wxcet},  
\cite{how02wxcet}), the 2001 and the 2004 ones \citep{ste07wxcet} for WX
Cet, the 2002 May and the 2006 ones for V844 Her \citep{oiz07v844her},
the 1992 one\citep{kat01hvvir} and the 2002 one \citep{ish03hvvir} for
HV Vir.
However, there is also a need for investigations on various superoutbursts
of each object such as SW UMa, since clear differences of the superhump
evolution have not been found.

\subsection{\it Humps After Superoutbursts}
\label{discussion3}

After the rapid decline phase of superoutbursts, some SU UMa-type dwarf
novae show periodic modulations with the ordinary superhump period, but
the phase is shifted by typically $\sim0.5$. This phenomenon is called
`late superhumps' (\cite{hae79lateSH}, \cite{vog83lateSH},
\cite{vanderwoe88lateSH}). 

In the light curves of the 2002 superoutburst of SW
UMa, the superhumps seem to have two peaks (the primary peaks, and the
secondary peaks are indicated 
by solid arrows and dashed arrows, respectively in figure \ref{02dailylc}).
At the beginning, the primary humps were prominent, and the secondary
humps progressively increased their amplitude at the end stage of the
superoutburst. Such growth of secondary humps were also observed in SW
UMa during the 1986 superoutbursts \citep{rob87swumaQPO}, the 1991 one
\citep{kat92swumasuperQPO}, the 1996 one \citep{sem97swuma}, and the
2000 one (this paper), and in addition, during superoutbursts of
other objects (e.g. \cite{sch80vwhyi}, \cite{uda90suuma}). \citet{sch80vwhyi}
and \citet{war95book} suggested that such secondary humps develop into
late superhumps. If secondary humps replace the primary during the
end stage of the superoutburst, and these humps have been regarded as late
superhumps, it is naturally explained that late
superhumps have shown the same period as that of the ordinary superhumps and
the shifted phase in previous works.

We observed humps on HJD 2451606, at the end of the rapid decline phase
of the 2000 superoutburst of SW UMa (figure \ref{00dailylc}).
Although no hump was clearly visible after HJD
2451606, we obtained a periodicity of 0.05798 days from the data
between HJD 2451607 and 2451610. This period is close to the superhump
period at the end stage of the outburst (0.05771 days), between HJD
2451603 and 2451606. However, no obvious phase shift of $\sim0.5$ was
observed (figure \ref{00latedailylc}). Humps after the rapid decline
phase of the 2000 superoutburst, therefore, seems not late
superhumps, but the remains of the ordinary superhumps.

\subsection{\it QPOs}

Quasi-Periodic Oscillations (QPOs) are observed in light curves of CVs
and in X-ray
binaries (see e.g. \cite{war08CVsQPO}). The typical amplitude and period
are $\sim0.01$ mag and $\sim300$ s, respectively, for CVs.
In SW UMa, \citet{rob87swumaQPO} observed QPOs which had a amplitude of
up to $\sim0.01$ mag and a period of $\sim4.8$ min during the 1986 superoutburst.
Following this, as mentioned in section \ref{introduction}, 
\citet{kat92swumasuperQPO}
discovered unusually large-amplitude QPOs ($\sim0.2$ mag) with a period
of $\sim6.1$ min during the 1992 superoutburst, and called them `super-QPOs'.
Although \citet{war03DNO3} also observed large QPOs with an amplitude of
$\sim0.2$ mag in quiescence of a dwarf nova, WX Hyi, such large QPOs are
very rare in CVs.  
During the 1996 superoutburst of SW UMa, \citet{nog98swuma} found
ordinary QPOs with an amplitude of $\sim0.01$ mag and a period of
$\sim5.3$ min, but they failed to detect so-called `super-QPOs'.

We detected QPOs with an amplitude of $\sim0.02$ mag at the end stage of
the 2000 and the 2002 superoutbursts (figure \ref{qpo}), but could not
find super-QPOs with such a large amplitude of $\sim0.2$ mag. These QPOs
we detected had 
a period of $\sim11$ min, which is about twice as long as those detected
at the early or middle stage of superoutbursts of SW UMa in previous
works. 
QPOs with a similar period detected at the similar phase of two outbursts
in our data indicate that these QPOs might have the same ogirin.

\section{CONCLUSION}
\label{conclusion}
Our main conclusion in this paper is summarized below:

\begin{enumerate}

\item We investigated superhump evolutions during the 2000, the 2002, and the
      2006 superoutbursts. After their appearance, the superhumps
      showed the same evolution during these superoutbursts, which
      implies that the superhump evolution may be governed by the
      invariable binary parameters, such as the mass ratio, the orbital
      period, and so on.

\item After the end of the 2000 superoutburst, we detected a periodicity
      close to the superhump in the light curve, but the phase shift
      which commonly accompanies late superhumps was not found. 

\item We found QPOs at the end stage of the 2000 and the 2002
      superoutbursts, but failed to detect so-called `super-QPOs' during
      three superoutbursts we investigated.

\end{enumerate}

\bigskip 

We are grateful to many observers who have reported vital
observations. We acknowledge with thanks the variable star
observations from VSNET, AAVSO, and AFOEV International Database
contributed by observers worldwide and used in this research. 

This work was supported by the Grant-in-Aid for the Global COE Program
"The Next Generation of Physics, Spun from Universality and Emergence"
from the Ministry of Education, Culture, Sports, Science and Technology
(MEXT) of Japan.

\end{document}